\documentstyle[12pt]{article}

\newcommand{\be}{\begin{equation}}
\newcommand{\ee}{\end{equation}}
\newcommand{\bn}{\begin{eqnarray}}
\newcommand{\en}{\end{eqnarray}}
\newcommand{\bd}{\begin{displaymath}}
\newcommand{\ed}{\end{displaymath}}

\begin{document}

\begin{flushright}
QMW-th/96-4
\end{flushright}

\begin{center}
{\Large \bf N=1 Heterotic/M-theory Duality and\newline Joyce Manifolds}
\newline
\newline
\newline
{\large B.S. Acharya}\footnote{e-mail: r.acharya@qmw.ac.uk. Work Supported
By PPARC}
{\it Queen Mary and Westfield College,}
{\it Mile End Road,}
{\it London. E1 4NS}

\end{center}

\begin{abstract}
We present an ansatz which enables us to
construct heterotic/M-theory dual pairs in
four dimensions. It is checked that this
ansatz reproduces previous results and that the
massless spectra of the proposed new dual pairs
agree.
The new dual pairs consist of $M$-theory
compactifications on Joyce manifolds of $G_2$
holonomy and Calabi-Yau compactifications of
heterotic strings. These results are further
evidence that $M$-theory is consistent on
orbifolds. Finally, we interpret these results
in terms of $M$-theory geometries which are
$K3$ fibrations and heterotic geometries
which are conjectured to be $T^3$ fibrations.
Even though the new dual pairs are
constructed as non-freely acting orbifolds of
existing dual pairs, the adiabatic argument is
apparently not violated.
\end{abstract}
\newpage
\section{Orbifolds and M-theory.}
The predictions of string dualities \cite{hull,witten} have given rise to
a fascinating web of interconnections between our most promising candidate
descriptions of nature. An underlying structure is slowly emerging. In
particular,
it has now become apparent \cite{witten,Hor} that the strong coupling
dynamics of both the TypeIIa and ${E_{8}{\times}E_{8}}$ heterotic
string theories can be understood in terms of certain one-dimensional
compactifications of $M$-theory. Specifically the TypeIIa theory
is related to $M$-theory on a circle, $S^1$, and the heterotic
string to $M$-theory on ${S^1}/{Z}_2$, where the $Z_2$ acts as
reflection on the coordinate of the $S^1$.

A precise definition of $M$-theory is yet to be made; however
consistency with the remarkable evidence that string theories
in various dimensions are connected, not only to each other
but to a supersymmetric theory in eleven dimensions ($M$-theory)
gives us some information about some of the properties of $M$-theory.
In particular, it seems clear that the low energy dynamics
of $M$-theory are described by eleven-dimensional supergravity
theory. Secondly, $M$-theory must share certain properties
with string theory. For example, $M$-theory on an orbifold
must be a consistent quantum theory \cite{Hor,Morb}. This
is certainly not a property shared by its low energy cousin.
Finally, $M$-theory must contain higher dimensional objects,
$p$-branes which play a fundamental role in the duality
conjectures. One viewpoint is that fundamental strings in
$d < 11$ arise from closed $p$-branes in $M$-theory, and
that Dirichlet-branes (D-branes) in string theory \cite{pol1} arise
from open $p$-branes in $M$-theory \cite{mbrane}.

One hope is that one may be able to derive {\it all} connections
between various string theories in lower dimensions from $M$-theory.

In \cite{witten} evidence was presented that the strong coupling
limit of the TypeIIa string theory in ten dimensions is effectively
described by eleven dimensional supergravity compactified on a circle.
The arguments leading to this conclusion are well known so we do not review
them here and instead concentate on the heterotic string.
In \cite{Hor} evidence was presented that $M$-theory on an ${S^1}/Z_2$
orbifold gives a description of the strongly coupled $E_{8}{\times}E_8$
heterotic string. The arguments were as follows:
The $Z_2$ kills one of the two supersymmetries present in $M$-theory
on a circle. The two fixed points which arise in the orbifold of the
theory define two fixed ten dimensional hyperplanes, on which
anomaly cancellation requires $E_{8}{\times}E_8$ gauge symmetry to be
present in the theory. This gauge symmetry was understood to have arisen 
from the twisted sectors of the orbifold, in analogy with string theory.

One is then led to the picture that $M$-theory on $X{\times}S^1$
is associated to the TypeIIa theory on $X$, with $X$ any space. Similarly
$M$-theory on $X{\times}{S^1}/Z_2$ is related to the $E_{8}{\times}E_8$
heterotic string on $X$. 

Consider then, $M$-theory on ${T^3}{\times}{S^1}/Z_2$. One expects that this
theory is related to the heterotic string on $T^3$. On the other hand, one
expects that the strong coupling limit of the heterotic string on $T^3$ is
related to eleven-dimensional supergravity on $K3$ \cite{witten}. This implies
various possibillities; one is
that whatever $M$-theory may be, on $T^{3}{\times}{S^1}/Z_2$ it is a theory
which at low energies looks like eleven dimensional supergravity on $K3$.
A second possibillity is that $M$-theory on two different spaces, $K3$
and $T^{3}{\times}{S^1}/Z_2$ are both related to the $T^3$ compactified
heterotic string\footnote{A related discussion of these compactifications
appeared in \cite{duff}, where $M$-theory on $K3{\times}{S^1}/{Z_2}$ was
considered. From one point of view this `should' give the heterotic string on
${T^3}{\times}{S^1}/{Z_2}$ which is inconsistent. It was therefore argued
that this $Z_2$ acts also on the $T^3$ giving the heterotic string on $K3$.}. 
We
will show that a strikingly similar result arises in lower dimensions for
dual $M$-theory/heterotic compactifications which have less supersymmetry.
Specifically, for $N=2$ and $N=1$ heterotic string compactifications
to four dimensions it will become apparent that there are again {\it two}
$M$-theory compactifications which arise as the duals of these heterotic
theories. 

The general ambition of this paper is to find a heterotic dual for 
compactifications of $M$-theory on various Joyce 7-manifolds \cite{J1,J2}.
We do this by means of an ansatz, which we present shortly. It will later
transpire that (at least locally), these Joyce manifolds are $K3$ fibrations.
It is thus natural to expect that, if the heterotic dual is the
correct one, then we are discussing a fibration of the seven-dimensional
duality between the $T^3$ compactified heterotic string and $M$-theory on
$K3$ by the fibrewise application of the seven dimensional duality in 
the adiabatic limit \cite{vaf}. 
However, if the heterotic dual is on some space $X$, then we also
expect that we have an $M$-theory dual compactification on 
$X{\times}{S^1}/Z_2$
\cite{Hor}. We can naturally interpret this as a fibration of the {\it other}
$M$-theory/heterotic duality in seven dimensions.

The aim of the remainder
of this section is to use an ansatz to rederive 
some of the previously constructed dual
pairs, where apart from the $Z_2$ orbifold which defines the $K3$, all
other elements of the orbifold group act freely. 

Consider then $M$-theory on $T^4$. We can take a $Z_2$ orbifold of this
theory in the following way.

\be
\alpha:(x_1,x_2,x_3,x_4) = (-x_1,-x_2,-x_3,-x_4)
\ee
where $(x_1....x_4)$ are the coordinates of the four torus.

This orbifold has sixteen fixed points and defines a particular orbifold
limit of $K3$. From our preceding discussions, one would expect that
$M$-theory on such an orbifold has certain twisted sectors, associated
with these fixed points, at which extra massless particles may arise.
Given the lack of a definition of $M$-theory it is difficult to make
any precise statements about such twisted sectors. However, one can
draw an analogy again with string theory whereby blowing up the
fixed points, we should recover the results of the theory on the resulting
smooth manifold. In this case, the smooth manifold is $K3$, and we
expect that in this limit, the theory is dual to the heterotic string
on $T^3$ \cite{witten}. 

We now come to the first and most crucial part of the ansatz which will
enable us to construct dual pairs of heterotic/$M$-theory compactifications
in lower dimensions. The {\it first} part of the ansatz is to label 
the heterotic $T^3$ coordinates
with the {\it same} labels as three of the $M$-theory $K3$ coordinates. We
will then toroidally compactify both theories to four dimensions on a further
$T^3$ with coordinates $(x_{5},x_{6},x_{7})$. We thus have $M$-theory on
$K3{\times}T^3$ and the heterotic string on $T^6$. The {\it second} 
part of the
ansatz is the following: we will take further $Z_2$ orbifolds of this $M$-
theory background giving vacua with $N=2$ and $N=1$ supersymmetry in four
dimensions. The isometries defining these orbifolds will act on the $M$-theory
coordinates, $(x_{1},...x_{7})$. The crucial point is that, because of the
labelling choice in the first part of the ansatz, the six coordinate labels of
the heterotic string on $T^6$ are a subset of the seven coordinate labels of
the $M$-theory compactification. Thus the definition of an orbifold 
isometry in the $M$-theory geometry {\it also} defines an orbifold isometry
in the heterotic geometry. In general, if we begin with $M$-theory on a 
$K3{\times}T^3$ orbifold defined by $\alpha$, as in eq.(1) 
and take a further orbifold
of the theory, generated by a group of isometries denoted by $\Theta$, then
because of the labelling choice in the first part of our ansatz, $M$-theory
on this ${T^7}/{(\alpha,\Theta)}$ orbifold should be equivalent to the
heterotic string on a ${T^6}/{\Theta}$ orbifold.

A priori this labelling choice
may seem like a rather bizarre thing to do; however, given this as an
ansatz and nothing more we will show that such a choice of labelling always
gives rise to the correct heterotic/$M$-theory spectra when we orbifold to
produce new dual pairs in lower dimensions. This only appears to work when
the only non-freely acting members of the orbifold group possess $K3$
orbifold singularities. More precisely, this construction only works when
the singular set of ${\Theta}$, consists solely of $K3$ orbifold 
singularities. For the rest of the paper we thus choose the heterotic 
$T^3$ coordinates as
$(x_1,x_2,x_3)$, where these are the {\it same} labels used to define the
$K3$ in $M$-theory. A clue that this is the correct choice to make was
given in \cite{duff} and we refer the reader there and to the previous 
footnote
for details. The fact that this strategy appears to work in all cases
strongly suggests that there is some as yet underlying structure to the way
one can construct $M$-theory/heterotic dual pairs in dimensions less than
seven. Further evidence of this has emerged in \cite{duff}.

Now let us toroidally compactify the $x_5,x_6$ and $x_7$ directions.
On $K3{\times}T^3$, we thus relate $M$-theory to the heterotic string
on $T^6$ and to the TypeII theory on $K3{\times}T^2$. This four dimensional
theory has $N=4$ supersymmetry.

We can further orbifold this theory by a $Z_2$ isometry which gives $N=2$
supersymmetry in four dimensions. This example was considered in \cite{ferr}.
The action of this isometry is defined on the seven coordinates on which
$M$-theory is compactified as follows:
\be
\beta:(x_1,.....x_7) = (-x_1 + 1/2,-x_2,x_3,x_4 + 1/2,-x_5,-x_6,x_7 + 1/2)
\ee
Because of the half shifts on the torus defined by $(x_{4},x_{7})$, 
this $Z_2$ acts
freely. When combined with the $Z_2$ defined by $\alpha$, the {\it blown up}
$Z_{2}{\times}Z_2$ orbifold will give $M$-theory on $CY_{11}{\times}S^1$;
where the Calabi-Yau manifold $CY_{11}$ is self-mirror and has $h_{11}$=11
\cite{ferr}.
This $M$-theory compactification is then related to the TypeIIa theory
on $CY_{11}$. This thus gives rise to an $N$=2 theory whose massless spectrum 
at generic points in the moduli space of the vector multiplets is $(12,12)$,
where following \cite{kach} $(M,N)$ denotes an $N$=2 theory with $N$
vector multiplets and $M$ hypermultiplets.

In \cite{ferr} and all known examples to date, the action of orbifold isometry
groups (which act on the TypeIIa theory on $K3{\times}T^2$) 
on the heterotic string on $T^6$
were calculated using the connection between the lattice of integral
cohomology
of $K3$ and the Narain lattice for the $T^4$ compactified heterotic string.
In other words, all existing dual pairs, constructed as orbifolds, have
been derived from string-string duality in six dimensions. 

Now, let us suppose that we know nothing about the cohomology of $K3$, but
that we know how to construct consistent heterotic string orbifolds.
Then we can ask how can we reproduce the $N=2$ $(12,12)$ spectrum from
a heterotic $Z_2$ orbifold? Because of our ansatz, the action of $\beta$ on
the $M$-theory geometry also defines its action on the $T^6$ of the heterotic
theory, because the $T^6$ coordinates we have chosen
are $(x_1,x_2,x_3,x_5,x_6,x_7)$. Thus, all that remains is to specify
the action on the gauge degrees of freedom ie the sixteen left movers.
Given that $\beta$ acts freely on $(x_3,x_7)$, we have an invariant
two-torus, which will give rise in general to four vector multiplets.
Thus, all that remains is to project out eight of the sixteen possible
additional vectors which are associated with the
Cartan subalgebra of the ten dimensional heterotic
gauge group. Because the orbifold is of
order two, we know that essentially the only possible action is exchanging
the two $E_8$ factors in the gauge group. Finally, in order to achieve
modular invariance we are {\it forced} to include an 
asymmetric $Z_2$ shift\footnote{Specifically, the vacuum energy 
in the left moving twisted sector is $-1/4$ which does not lead
to a modular invariant orbifold. However if we translate the
shift on $x_7$ in the $M$-theory background
to an asymmetric shift of the $x_7$ in the heterotic
background, then we can achieve modular invariance in the
following way \cite{ferr}: the ${\Gamma}^{1,1}$ which corresponds
to $x_7$ can be orbifolded by a shift vector $\delta$ of
the form $\delta$ = $(p_{l},p_{r})/2$ with $p^2$ = $2$. Because
${{\delta}^2}/2$ = $1/4$, the difference between left and right moving
vacuum energies is zero, and hence the orbifold
is modular invariant.}which is related to the shift on $x_7$.
This reproduces the model of \cite{ferr} without any
knowledge of the cohomology of $K3$.

In a similar manner, one can consider a further freely acting $Z_2$
orbifold of the above $N=2$ theories which was considered in \cite{har}
to produce dual theories with $N=1$ supersymmetry in four dimensions.
Again without using any knowledge about the cohomology of $K3$ one can
reproduce the result of \cite{har}. We, of course, are not suggesting
that the identification of the lattice of integral cohomology of $K3$ with
the Narain lattice for the heterotic string on $T^4$ is incorrect. 
For the examples 
considered in \cite{har,ferr}, which we reproduced above, the $K3$ orbifold
defined by $\alpha$ in equation (1) is the only element of the orbifold
group of this type. We make the supposition that we know nothing of the
cohomology of $K3$, in order to proceed further and construct dual pairs
when two or more elements of the orbifold group are of the $K3$ orbifold type.
The identification between these lattices will be seen to hold in the 
adiabatic limit \cite{vaf} when, in section four, we interpret our results in
terms of the `fibration picture' of \cite{vaf}.

In the examples we have just considered we resolved {\it all} singularities
because it is unclear at present how to deal with twisted sectors in
$M$-theory. Luckily there were no singularities associated with
$\beta$, so there were none to resolve. However in the more general
cases we will consider, we will orbifold the heterotic theory with
isometries that do have singularities and it is natural to resolve ie
blow up these as well. The blowing up modes may naturally be associated
with the twisted sectors of the heterotic theory.

This suggests the following strategy:
(I) Take $M$-theory on an orbifold. By analogy
with string theory, we can naturally identify the twisted sector states
with the blowing up modes of the orbifold.
(II) The action on the $T^6$ of the heterotic string will already be
specified by the action of the orbifolds on the geometry of $M$-theory
by our ansatz.
(III) Then simply project out the necessary number of vectors from the
heterotic string spectrum.
(IV) If the heterotic orbifold is a $(2,2)$ superconformal field theory, in
which case the blowing up modes are truly moduli \cite{dix} then proceed
to the smooth limit and include the blowing up moduli in the spectrum.
In this case one must choose an embedding of the spin connection
in the gauge connection such that the resulting spectrum is correct.
In fact, if the orbifold group acts left-right symmetrically on the
${\Gamma}^{6,6}$ of the $T^6$ compactified heterotic string, then
the orbifold has a classical geometric interpretation and one
can study the heterotic string on the blown up orbifold and the
moduli of the smooth manifold will in any case appear as scalar fields
of the theory. Because such a theory is a string theory on a blown up
orbifold, the massless spectrum is easy to determine: it is just
the untwisted sector at the orbifold limit plus the moduli associated
with blowing up.
(V) Check if the theory is consistent with modular invariance or can be
made so by adding appropriate shift vectors.

In fact, it may be possible to go further than just considering the theories
on blown up orbifolds. We will give strong evidence in some four
dimensional $N=1$ examples that both the untwisted and twisted sector spectra
coincide for $M$-theory on Joyce orbifolds and the heterotic string on
Calabi-Yau orbifolds. In the following sections, we will apply our
presented strategy to propose new dual pairs. These constitute
examples of dual pairs constructed as non-freely acting, supersymmetry
breaking(ie less than $N=4$ in $4d$) orbifolds of existing
dual pairs. Section two discusses an $N=2$ example in detail. In
section three we construct some $N=1$ examples. In section four we interpret
our results in terms of the fibration picture in the adiabatic 
limit \cite{vaf}. Remarkably, even though our examples are constructed as 
non-freely acting orbifolds of existing dual pairs, the adiabatic argument
of \cite{vaf} is apparently not violated. This is a consequence of the
fact that the orbifolds we restrict ourselves to are precisely the ones
which preserve the fibration structure. Following this we
end with some conclusions and comments.

\section{An $N=2$ Example.}
In this section we will consider an $N=2$ example first following
the analysis given in \cite{kach} and then following the
strategy presented in the last section.
In \cite{kach} several examples of potential dual pairs
of $N=2$ theories in four dimensions
were constructed. The dual pairs in question were Calabi-Yau
compactifications of TypeII strings and $K3{\times}T^2$
compactifications of heterotic strings. The massless spectrum of the 
heterotic string on $K3{\times}T^2$ is determined from the expectation value
of the gauge fields on $K3$ and index
theory \cite{kach,green}. An $N=2$ theory in four dimensions
is characterised by vector multiplets
and hypermultiplets. The vector multiplets contain adjoint scalar fields,
which in addition to the moduli hypermultiplets of the theory are also 
moduli. At special points in the moduli space of these
scalars, the theory contains massless charged hypermultiplets.
These become massive at generic points in the moduli space of the
adjoint scalars which correspond to the Cartan subalgebra of the
gauge group. Neutral massless fields will always remain massless
as one moves through the moduli space of these scalars. After
the spin connection has been embedded in the gauge connection
in such a way that the theory is anomaly free,
the theory is then characterised by the neutral fields (the
rest of the moduli) and $N$ vector multiplets, where $N$ is the rank of the
gauge group which survives the embedding. Kachru and Vafa denoted the
theories at these generic points\footnote{ 
We assume we are at generic points in the torus.}
by $(M,N)$, where $M$ is the number
of massless hypermultiplets.
There is a universal contribution of $20$ to $M$ coming from the moduli
of $K3$ \cite{pk}, hence
$M\geq20$. Further, because we have a rank sixteen gauge
group in ten dimensions and a further four $U(1)$'s coming from the torus
at generic points, $N\leq20$.

We wish to consider an example where both
these inequalities are saturated. This places two constraints on possible
embeddings of the gauge bundle on $K3$: {\it (i)} we are forced
to consider giving expectation values to $U(1)$ or products of $U(1)$
gauge fields on $K3$. Fortunately, the spectra of many of these embeddings
has been calculated in \cite{green}, although we know not of any 
full classification; {\it (ii)} in order that we obtain $20$ moduli
hypermultiplets we need to find an example in which all hypermultiplets
are charged except the $20$ gravitational moduli.

Examining the spectra given in \cite{green}, it is not too
difficult to convince oneself that many of the examples satisfy these
criteria, giving a $(20,20)$ spectrum at generic points. For definiteness
we consider the model with a single $U(1)$ embedded in $E_{8}{\times}E_{8}$,
in such a way that one $E_8$ is broken to $E_{7}{\times}U(1)$. This gives
a spectrum containing 10 ${\bf 56}$'s of $E_7$(half with one
$U(1)$ charge and half with the opposite charge) 
and 46 $E_7$ singlets(23 each with opposite $U(1)$ charges).
This is the example given in equation {\it (6.1)} of \cite{green}.
Because all matter is charged, we have a $(20,20)$ spectrum at generic
points.

As an aside, it is interesting to note that
we can {\it connect} this example to the
chain of examples considered in section 3 of \cite{kach}. Namely, by
higgsing the $U(1)$, we find $45$ additional gauge neutral hypermultiplets
giving a $(65,19)$ spectrum. This model can be similarly higgsed \cite{kach}
several times to give the chain: $(20,20) \rightarrow (65,19)
 \rightarrow (84,18) \rightarrow (101,17) \rightarrow (116,16)$.
 \footnote{While this work was in progress, we realised
 that many of the
 examples of \cite{green} are connected via Higgs' and Coulomb branches
 and which have TypeII dual candidates on $K3$-fibrations \cite{klemm}.}

We would now like to find a TypeII dual for this $(20,20)$ model.
According to \cite{k}, the Calabi-Yau space describing the background
of the TypeII theory should be a $K3$ fibration. Further it must
be a self-mirror Calabi-Yau with $h_{11}$=$h_{21}$=$19$ because
a $(M,N)$ model arises from a TypeIIA compactification on a
Calabi-Yau with $(M=h_{11}+1,N=h_{21}+1)$. Luckily there is a
manifold which fulfills these requirements. We denote this Calabi-Yau
by $CY_{19}$. It can be constructed as a blown up orbifold as we now
describe, following Joyce \cite{J2}.

In \cite{J2},
Joyce constructed ${CY}_{19}{\times}{S^1}$ 
as a ${Z}_{2}{\times}{Z}_{2}$ blown-up orbifold of
the seven torus. We repeat the construction here:

 Define the seven-torus coordinates as $(x_1,......,x_7)$.
 Two $Z_2$ isometries of $T^7$ are defined by:
 \be
 \alpha(x_1,....x_7) = (-x_1,-x_2,-x_3,-x_4,x_5,x_6,x_7)
 \ee
 \be
\beta(x_1,....x_7) = (-x_1,1/2-x_2,x_3,x_4,-x_5,-x_6,x_7)
 \ee
 Let the $Z_2{\times}Z_2$ isometry group generated by $\alpha$ and
 $\beta$ be denoted by $\Gamma$.
In fact it is easy to see that this is precisely the construction
of \cite{ferr}, without the extra $Z_2$ shifts which made that
construction freely acting.
 If considered separately, each of these $Z_{2}$'s has 16 fixed $T^{3}$'s
 and each one defines an orbifold limit of a particular $K3{\times}T^{3}$. 
 Hence, the singular set\footnote{In general we define the singular 
 set $S^{\prime}$ of $M$ to be the set of points, surfaces and submanifolds 
 of the manifold $M$, which are fixed under the action of some finite 
 group $G$. The singular set, $S$, of $M/G$ is then the image of $S^{\prime}$
 in $M/G$.}
 of $T^{7}/{\alpha}$ contains 16 ${T^3}$
 components, as does the singular set of $T^{7}/{\beta}$. 
 However, as $\beta$ acts freely on the 16 fixed three tori of $\alpha$,
 $\alpha$ contributes eight three tori to the singular set
 of $T^{7}/{\Gamma}$. Similarly, $\beta$ also contributes eight three tori
 to the singular set of $T^{7}/{\Gamma}$.

The betti numbers of the original torus which survive the orbifold
projection ie the betti numbers of $T^{7}/{\Gamma}$ 
are $b_{1} =1$, $b_{2} =3$ and $b_{3} =11$. The blowing up
procedure is carried out by inserting non-compact Eguchi-Hanson 
geometries(${\times}T^{3}$)
in each of the singular regions. Each of these adds $1$ to $b_{2}$ and $3$ to
$b_{3}$, giving a seven manifold of $SU(3){\times}1$ holonomy with betti 
numbers: $b_{1} =1$, $b_{2} =19$ and $b_{3} =59$. In particular,
if we consider the six-torus defined by the coordinates $x_1$ through $x_6$
, then the holomorphic three form is preserved by the $Z_{2}{\times}Z_{2}$.
The seven manifold thus
obtained has the form  ${CY}_{19}{\times}{S^1}$. Compactification of eleven
dimensional supergravity on this manifold yields $N=2$ supergravity
with $20$ hypermultiplets and $19$ vector multiplets (not including the
graviphoton). The counting goes as follows
\footnote{See \cite{KK} for a review}: in eleven dimensions,
the massless bosonic fields of $M$-theory are the
metric, ${G}^{{\mu}{\nu}}$, and antisymmetric three-form tensor,
${A}^{{\mu}{\nu}{\rho}}$.
On compactification to four dimensions on $CY_{19}{\times}S^1$, the three
form gives rise to $b_{3}$ scalars, $b_{2}$ vectors and $b_{1}$ two forms
\footnote{In four dimensions, two forms are dual to scalars and so may
be counted as scalars.}. In general a higher dimensional metric yields
$n$ scalars, where $n$ is the dimension of the moduli space of the
compactification metric. In our case, this is $59 - 1 =58$. The metric
tensor will also yield a vector in the lower dimensional theory for
every continuous isometry of the compactifying manifold. In our case,
the $S^1$ has a $U(1)$ isometry yielding a $U(1)$ gauge field in
four dimensions. 

All in all, for this example we get $118$ scalars and $20$ vectors 
(including graviphoton) plus the graviton. The fermion spectrum is 
implied by supersymmetry and we thus have the $(20,20)$ model as
required. This is the same spectrum as the Type IIa/IIb
string on $CY_{19}$. 

Now let us apply the strategy suggested in the last section and
see if it gives the correct results. 
Firstly, the action of $\beta$ on the heterotic string $T^6$
does indeed give an orbifold limit of $K3{\times}T^2$ as before.
Now, however, we do not have the extra half shift on $x_7$, so this
orbifold is not freely acting. Secondly, we would like a rank 20
gauge group, which means that the action on the gauge degrees of freedom is
trivial. But, this is not the whole story, for we would certainly
like to preserve modular invariance in this orbifold and 
the natural choice we make is the standard embedding.
Away from the orbifold limit ie on the smooth $K3{\times}T^2$ we must
also specify an anomaly free background; and, as mentioned at the beginning
of this section this will limit us to $U(1)$ embeddings of the
spin connection in the gauge connection to give the required $(20,20)$
spectrum as before. Thus it appears that our strategy is consistent
at least for the first example we have considered.

We can provide a further check on whether we have indeed produced a dual pair,
by considering a freely acting orbifold of this dual pair. If the spectra
again agree, we will have also produced another dual pair. A simple freely
acting orbifold which does not break any supersymmetry is the following:
\be 
\sigma(x_1,x_2,...x_7) = (x_1,x_2+1/2,x_3,x_4,x_5,x_6,x_7)
\ee
On the M-theory geometry, this isometry has the effect of
halving the number of elements of the singular set of ${T^7}/{\Gamma}$. This
produces an example with a $(12,12)$ spectrum at generic points.

On the heterotic side, the action of $\sigma$ corresponds to exchanging the 
two ${E_8}$ factors of the gauge group, plus identifying, in
eight pairs of two, the sixteen fixed points associated with $\beta$ which
defines the $K3$ orbifold heterotic background. The shift on $x_2$ also 
eliminates massless modes coming from the $\sigma$ twisted sector. In
fact the twist by $\sigma$ is precisely the one which was considered in 
\cite{chl}.
Because the sixteen fixed points of $\beta$ on the heterotic background
are associated with sixteen neutral moduli hypermultiplets in the blowing up
limit, the $\sigma$ action reduces this number to eight. The resulting
spectrum is therefore precisely $(12,12)$ at generic points in accord
with the expectations of string-string duality.

As a concluding remark to this section, it is useful to point out that
because the respective 
moduli spaces of these conjectured dual pair of compactifications
are constrained by $N=2$ supersymmetry, many of 
the important results of \cite{ferr}
also apply here.

\section{$N=1$ Examples.}
Compactification of $M$-theory on a seven-manifold of
$G_2$ holonomy gives rise to $N=1$ supergravity with $b_2$ vector multiplets
and $b_3$ chiral multiplets. We denote these manifolds by ${J}^{b_2}_{b_3}$.
Many examples of such manifolds were recently constructed in \cite{J1,J2}.
A duality between the eleven dimensional theory on ${J}^{16}_{39}$ and the
heterotic string on a Calabi-Yau with precisely the same Hodge diamond as
${CY}_{19}$ was conjectured in \cite{pap}, on the basis of counting
Betti numbers and matching the spectra. We will see later in this section
that we can {\it derive} this result utilising our presented ansatz.

In this section, we proceed to apply the strategy of the preceeding sections
to produce dual pairs with $N=1$ supersymmetry in four dimensions. The first
example we consider has $b_2$=$8$ and $b_3$=$31$.\footnote{To the best of our
knowledge, this Joyce manifold has not been constructed previously, even
though a manifold with the same betti numbers appeared in \cite{J2}.} This
example will be constructed by considering a freely acting orbifold of the
$(20,20)$ $N=2$ example of the previous section. Evidence for the existence
of the $N=1$ dual pair, is then also evidence for the $N=2$ dual pair.

Consider then the $\Gamma \equiv $${Z_2}^3$ orbifold of the seven torus 
defined by $\alpha$
and $\beta$ of equations $(3)$ and $(4)$; and the third $Z_2$ defined as
follows:
\be
\gamma(x_1,x_2,.....x_7) = (1/2-x_1,x_2+1/2,-x_3,x_4,1/2-x_5,x_6,-x_7)
\ee
Because of the half shift on $x_2$, $\gamma$ acts freely. In fact, $\gamma$
takes the sixteen elements of the singular set of ${T^7}/{(\alpha,\beta)}$
and identifies them in eight pairs of two. The betti numbers of
${T^7}/{\Gamma}$
are $b_2$=$0$ and $b_3$=$7$. The singular set contains eight elements, the
resolution of each of which adds $1$ to $b_2$ and $3$ to $b_3$, giving a
Joyce manifold of $G_2$ holonomy, ${J}^{8}_{31}$.

Thus far, we have said little about the possible gauge groups allowed by
string/string/M-theory duality. The mechanism for gauge symmetry enhancement
in the TypeII theories is a generalisation of that considered in \cite{str}, 
where p-brane solitons
wrap around p-cycles of the compactification space and give rise
to massless gauge multiplets (and matter multiplets) when the cycles
degenerate to zero volume. In general, because the singularities
corresponding to the vanishing cycles are of A-D-E type, one expects
A-D-E symmetries \cite{witten,paul}. 
The singularities we have been considering
are all $SU(2)$ orbifold singularities, thus we can at least expect
an $SU(2)$ factor in the gauge group for each element of the singular
set that we blow up. For example, in the $N=2$ $(20,20)$ example
that we constructed, we should expect an ${SU(2)}^{16}$ factor in the
gauge group if we consider $M$-theory at the orbifold
limit defined in the previous section, 
because we resolved sixteen $SU(2)$ singularities on
the TypeII side to construct $CY_{19}$. So let us break the
${E_8}{\times}E_{8}$
gauge symmetry of the heterotic string to ${SU(2)}^{16}$ by
orbifolding the ${\Gamma}^{22,6}$ Narain lattice for toroidal
compactification to four dimensions. This is equivalent to the turning on of
Wilson lines. We will orbifold the theory by three $Z_2$ shift vectors
given by:
\newline
${\delta}_{1} = (1,{0}^{7};1,{0}^{7};1/2,{0}^{5})(1/2,{0}^{5})$;
\newline
${\delta}_{2} = ((1/2)^{4},{0}^4;(1/2)^{4},(0)^{4};0,1/2,(0)^{4})(0,1/2,(0)^{4
})$ 
\newline
and
\newline
${\delta}_{3} = ((1/2)^{2},(0)^{2},(1/2)^{2},(0)^{2};
(1/2)^{2},(0)^{2},(1/2)^{2},(0)^{2};(0)^{2},1/2,(0)^{3})$
\newline$((0)^{2},1/2,(0)^{3})$. 
\newline
This then breaks the ${E_8}{\times}E_{8}$ symmetry to ${SU(2)}^{16}$ as
required.\footnote{This choice of symmetry breaking vectors was considered 
in \cite{har}.} The action of $\beta$ from equation (4) on the heterotic
string is as in the given previous section, but now $\beta$ acts on the
theory with reduced symmetry. 

Because each $SU(2)$ factor in the gauge group is associated with an
orbifold singularity on the TypeII/M-theory side of the duality map,
the action of $\gamma$ on the heterotic string gauge group
is easily seen from the discussion above to be the exchange of the
two ${SU(2)}^{8}$ factors. The following action on the $T^6$ coordinates is
given by the ansatz of the preceeding sections.
\be
\gamma(x_1.....x_6) = (1/2-x_1,x_2+1/2,-x_3,1/2-x_5,x_6,-x_7)
\ee
In the untwisted sector of the theory the massless spectrum is
given by eight $N=1$ vector multiplets and $15$ chiral multiplets. Of
the chiral multiplets, seven are singlet untwisted moduli multiplets
and the other eight are adjoint multiplets of $SU(2)$. The $\gamma$ and
$\beta\gamma$ twisted sectors produce no massless states. The $\beta$
twisted sector produces $64$ chiral multiplet $SU(2)$ doublets, of
which only $16$ are $\gamma$ invariant. Hence the resulting
massless spectrum is precisely that of $M$-theory on ${J}^{8}_{31}$.

We can modify this example slightly and produce nine
more potential $N=1$ dual examples. This is done as follows:
\be
\gamma(x_1,....x_7) = (1/2-x_1,x_2,-x_3,x_4,-x_5,x_6,-x_7)
\ee
This modification has the following significance: 
{\it (i)}: $\gamma$ is no longer freely acting on the geometry;
{\it (ii)}:now the element
$\alpha\beta$ acts trivially on the fixed three tori of $\gamma$. This
means that the presence of $\gamma$ in this form removes four elements of
the singular set of $\alpha$ and $\beta$ of the original $N=2$ model.
So we definitely have eight vectors surviving
the $\gamma$ projection. However, we still need to consider the elements
of the singular set induced by $\gamma$. Because the element $\alpha\beta $
acts trivially on the singular set from $\gamma$, $\gamma$ must contribute
eight additional elements. However, these elements are different to the
the other eight, because when the blowup is performed, the additional 
action of $\alpha\beta$ must be considered on the blowup itself. It
turns out that there are two topologically distinct ways of considering
this action on the blowing up modes. These two ways differ by the fact that
one preserves the generator of $H^{2}(X,R)$ and the other changes its sign,
(where $X$ is the Eguchi-Hanson blowing up mode)\footnote{\cite{J1,J2}
can be consulted for further details.}.

It follows that these two blowups contribute different betti numbers to
the Joyce manifold. If the extra $Z_2$ action was not present then each
blowup would add one to $b_2$ and three to $b_3$. When the $Z_2$ is present,
the two choices in defining its action on the blowup 
has the effect of splitting these original betti numbers, 
so that the first type of resolution adds
one to $b_2$ and one to $b_3$; and the second adds zero to $b_2$ and
two to $b_3$. So all in all we have eight `standard' blowups and
eight for which there are two choices. The betti numbers from the original
seven torus which survive the $Z_{2}^{3}$ isometries are $b_1$=$b_2$=$0$ and
$b_3$=$7$. The eight  `standard' blowups add one to $b_2$ and three
to $b_3$, giving $b_{2} = 8$ and $b_{3} = 31$. Of the remaining eight
`nonstandard' blowups, if we choose $l$ of them to be of the first
type, then this adds $l$ to both $b_2$ and $b_3$. The remaining $8-l$ add
zero to $b_2$ and $16-2l$ to $b_3$. This means that the Joyce manifold
has
\be
b_{2} = 8 + l   ,  b_{3} = 47 - l   ,  l=0,1,...8
\ee
Let us now see if we can find a family of heterotic duals for this family
of Joyce manifolds.

The first point to note is that if we consider the ${Z_2}{\times}{Z_2}$
orbifold of $M$-theory on $T^7$ which is generated by $(\alpha,\gamma)$
then the resulting $N=2$ spectrum is precisely $(20,20)$ ie the
manifold is of the form ${CY_{19}}{\times}S^{1}$. This is the same
spectrum we
obtained using $(\alpha,\beta)$ as the orbifold generators. Hence,
the action of $\gamma$ is identical to that of $\beta$. This symmetry
between the generators should be preserved when we consider the action
of $\beta$ and $\gamma$ on the heterotic theory. Because we have already
found that the action of $\beta$ preserved the rank of the gauge
group originating in ${E_8}{\times}{E_8}$ in our $(20,20)$ $N=2$ model, 
we expect $\gamma$ to do so also. \footnote{Note, for the reasons
stated above, we again will consider the ${E_8}{\times}{E_8}$ broken
down to ${SU(2)}^{16}$.} Further, because the heterotic model will have 
$N=1$ supersymmetry in four dimensions, we can expect a rank 16 gauge group.
We therefore may expect on these general grounds that the heterotic model
will be dual to $M$-theory on the Joyce manifold with $l=8$, above, which
has a spectrum of $16$ vector multiplets and $39$ chiral multiplets. We
now construct the heterotic background.

Before considering the action on the gauge degrees of freedom, we
first specify the action of $\beta$ and $\gamma$ on the six-torus
coordinates of the heterotic string. According to our ansatz, these
are as follows:
\be
\beta(x_1,..x_6) = (-x_1,1/2-x_2,x_3,-x_5,-x_6,x_7)
\ee
\be
\gamma(x_1,...x_6) = (1/2-x_1,x_2,-x_3,-x_5,x_6,-x_7)
\ee

$\beta$ leaves invariant a two-torus which is inverted by $\gamma$; and
$\gamma$ leaves invariant a two-torus inverted by $\beta$. Thus
the heterotic background has $N=1$ spacetime supersymmetry as expected
by duality. In fact it is interesting to note that if this orbifold
is blown up, the resulting smooth manifold is none other than the ${CY}_{19}$
that appeared on the TypeII side in our $N=2$ example. If we now
consider the heterotic theory on the manifold ${CY}_{19}$ we will
find a non-chiral $N=1$ theory with $39$ moduli multiplets. If we also
specify a ${U(1)}^n$ embedding of the spin connection in the gauge connection,
then we arrive at the rank sixteen model conjectured previously in \cite{pap}!

However, we can go one stage further and actually give evidence that $M$-theory
on the orbifold defined by $({\alpha,\beta,\gamma})$ is equivalent to the
heterotic string on the orbifold defined by $({\beta,\gamma})$. The action
of the orbifold on the $T^6$ piece of the heterotic theory has been given.
It remains to specify the action on the gauge degrees of freedom.
As already noted, we wish to preserve a symmetry between $\beta$ and $\gamma$,
ie they should give rise to identical spectra when considered separately.
This can be achieved by considering identical embeddings of the
spin connection in the gauge connection, with the $\beta$ connection
in one ${SU(2)}^8$ factor and the $\gamma$ connection in the other.
The choices are restricted by modular invariance. Further, because we
expect $32$ chiral multiplets in $M$-theory to arise from the twisted sectors
{\it (because 32 harmonic three-forms arise from blowing up)}, we can
expect the same in the heterotic theory.
We find there are essentially two inequivalent
choices of abelian embeddings which give rise
to massless states in the twisted sector. Only one choice, corresponding to
the standard embedding in each ${SU(2)}^8$ factor,
gives rise to the correct number of twisted sector multiplets. These are
the following:
\be
{\delta}_{\beta} = ((1/2)^{2},(0)^{6})((0)^{8})
\ee
\be
{\delta}_{\gamma} = ((0)^{8})((1/2)^{2},(0)^{8})
\ee
where the first(second) bracket denotes the shift in the first(second)
${SU(2)}^{8}$ factor. Let us consider the spectrum.
In the untwisted sector, the spectrum contains $16$ $N=1$ vector
multiplets and seven moduli chiral multiplets (including dilaton).
In fact, with the ansatz we have alluded to one will {\it always} find
seven moduli multiplets in the untwisted sector. The analogue of
this statement from the $M$-theory point of view is that {\it any}
Joyce manifold of $G_2$ holonomy constructed as a blown up ${Z_2}^3$
orbifold of the seven torus has $b_{3}({T^7}/{{Z_2}^3})$=$7$, corresponding
to seven chiral moduli multiplets in the untwisted sector of $M$-theory!

Now consider the twisted sectors. We find $16$ $SU(2)$ doublet multiplets
from each of the $\beta$ and $\gamma$ sectors respectively. This gives
a total spectrum of $16$ vector multiplets and $39$ chiral multiplets.
This is the same spectrum as the example with $l=8$ above, as we initially
expected. Before commenting on the examples with $l=0,1..7$ we would
like to make some observations.

Firstly, if we examine the Calabi-Yau orbifold of the heterotic theory
defined by equations (13),(14), we note that if we resolved all
the orbifold singularities then the resulting Calabi-Yau manifold is
none other than $CY_{19}$! We have thus derived the result presented in
\cite{pap}.

Secondly,
we have seen that the untwisted matter content will {\it always} agree for
heterotic/$M$-theory duals if the $M$-theory background is a ${(Z_2)}^3$
orbifold of $T^7$. In the example we have just considered, we have further
observed that the twisted sector spectrum in the heterotic theory
precisely reproduces the spectrum which arises in $M$-theory from the
blowing up procedure. This is compelling evidence that we have again
constructed the correct heterotic background, dual to $M$-theory on
a Joyce manifold. It is also further evidence that orbifold backgrounds
are consistent in $M$-theory. In fact similar reasoning also applies to 
$N=2$ dual pairs. The $M$-theory background for $N=2$ supersymmetry in
four dimensions will be of the form $CY{\times}S^1$, for $CY$ any
Calabi-Yau space. This background is equivalent to the
TypeIIa theory on $CY$. If $CY$ is constructed as a ${Z_2}{\times}{Z_2}$
orbifold of $T^6$, then the untwisted matter spectrum at generic points
will {\it always} contain four massless hypermultiplets. The heterotic
dual background will then be of the form ${{T^4}/{Z_2}}{\times}T^2$ where the
$Z_2$ defines a $K3$ orbifold. This background also contains four
hypermultiplets in the untwisted sector massless spectrum, and it is yet
again tempting to postulate that the twisted sector of $M$-theory on
the ${(Z_{2})}^2$ orbifold is identical to that of the heterotic $Z_2$
orbifold. Of course, the heterotic spectrum depends strongly on
the choice of discrete Wilson lines or shift vectors required for
modular invariance and it would be interesting to identify such degrees
of freedom in $M$-theory. Such an identification was made for the TypeII
theory recently \cite{N2} where it took the form of generalised discrete
torsion.

We have identified a heterotic dual theory for the compactification
of $M$-theory on one of a family of nine Joyce manifolds, parametrised
by $l$. What can we say about the other members of the family?

Consider first $M$-theory on the example with $l=8$ ie ${J}^{16}_{39}$. 
To make the
transition to the next member of the family, ${J}^{15}_{40}$, we must
blow down a two-cycle and blow up a three-cycle; in other words
this is precisely an example of an $M$-theory conifold type transition, and
we have reasonable grounds to suspect that such a transition is physically
non-singular \cite{str}. In fact this is nothing but the Higgs mechanism
in an $N=1$ $M$-theory background. From the heterotic string point of view,
we would need a field content which along certain Higgs directions reproduces
the field content of all these Joyce compactifications of $M$-theory.
However, we have chosen the most simple breaking of ${E_8}{\times}{E_8}$
and have restricted ourselves to abelian embeddings of the spin connection
in the gauge connection. There are of course many other consistent 
possibillities that one may consider and it may certainly be the case that
we can reproduce the required field content from the heterotic 
compactification. We are investigating such possibillities. Thus the
heterotic analogue of these topological transitions between different Joyce
manifolds remains a mystery. This is also a consequence of the fact that the
singularities which allow transitions between the different Joyce manifolds
are certainly not $SU(2)$ singularities, but orbifolds of them. One
would certainly need to identify the physical implications of this statement
from the $M$-theory point of view, before the heterotic description of the
transition could be made.

\section{Interpretation.}
Given the apparent success of our ansatz, the question arises as to whether
these results have a more satisfying explanation. It is natural to expect that
this should come from some substructure in the Joyce manifolds considered in
this paper. A clue comes from the ubiquity of $K3$ fibrations in string 
duality \cite{klemm,k,vaf,M}. Specifically, given a $K3$ for the five brane
of $M$-theory to wrap around and an $S^1$ for the dual two-brane to wrap
around, one can derive connections between string 
theories in lower dimensions and $M$-theory by fibering the $K3$ over
another space \cite{k,vaf,M}. For example $N=2$ string-string
duality in four dimensions has an interpretation in terms of $N=2$ string-
string duality in six dimensions, whereby the geometries of the TypeIIa
theory ($K3$) and heterotic theory ($T^4$) respectively are fibered over
$CP^1$ \cite{vaf,k}. We will now show that (at least locally) a similar
interpretation holds here for the results of the preceding sections. More
precisely, if the $M$-theory compactification is a $K3$ fibration over some
three manifold and the heterotic dual compactification is a $T^3$ fibration
over the {\it same} three manifold, then in the adiabatic approximation of
\cite{vaf}, we can expect the duality to hold between the two theories
in four dimensions. The
following analysis relies heavily on that of \cite{J1,J2}.

A seven manifold of $G_2$ holonomy has two classes of special submanifolds.
This is essentially a consequence of the fact that the torsion-free $G_2$
structure of a Joyce seven-manifold is defined by specifying a particular
closed three form \cite{J1,J2}. This is analagous to the Kahler form in
Calabi-Yau spaces. The three form has a four form Hodge dual. The three
form is naturally identified with the volume form  of a special class
of three manifolds which are submanifolds of the Joyce manifold. These are
known as associative submanifolds. They minimize volume in their
homology class. The three cycle dual to the (minimum) volume form is known
as a supersymmetric cycle \cite{beck}. 

Similarly the four form which is dual to the three form can be used to define
special four dimensional submanifolds of the Joyce manifold. These are known
as coassociative submanifolds. It was proven in \cite{mac} that the dimension
of the moduli space of such manifolds is given by the number of self-dual,
harmonic two-forms of the submanifold. Therefore, if this number is non-zero,
then the Joyce manifold is (at least locally) fibered by the coassociative
submanifold.

Joyce \cite{J2} has given a prescription for proving whether or not such
submanifolds exist for the manifolds he constructs. The argument proceeds in
the following way: given a Joyce manifold, one can consider further orbifolds
by isometries which preserve the holonomy structure. If the singular set of
the isometry consists of $n$ elements of dimension three (eg three-tori) then
the Joyce manifold contains $n$ associative submanifolds (given by these three
-tori). If the singular
set contains $n$ dimension four elements (eg $K3$) then these are 
coassociative submanifolds of the Joyce manifold.

If we now begin with the duality in seven dimensions between $M$-theory on
$K3$ and the heterotic string on $T^3$, which was our original starting point,
then we would expect that fibering both of these manifolds over the {\it same}
three manifold will lead to a duality in four dimensions between these two
theories, at least in the adiabatic approximation \cite{vaf}. We will show,
following Joyce \cite{J1,J2}, that the seven manifolds used for $M$-theory
compactification in this paper are (at least locally) fibred by $K3$. Further,
we will similarly show that the seven manifolds contain $T^3$ submanifolds,
which ``pull back'' to the six manifold used for the heterotic 
compactification.
Unfortunately, it is not yet known whether or not these three-tori fiber the
$M$-theory and heterotic geometries, so we cannot say conclusively that the
fibration picture \cite{vaf} holds. However, given our preceding evidence
for dual pairs in four dimensions it is natural to expect that the $M$-theory
(heterotic) compactification space is globally a $K3$ ($T^3$) fibration.

Let us consider as an example our $N=2$, $(20,20)$ model of section 2, which
is defined on the $M$-theory background by equations $(3)$ and $(4)$. This
gave us an $M$-theory compactification on ${CY}_{19}\times{S^1}$. Consider
now the following isometry of this manifold:
\be
\sigma(x_1....x_7) = (x_1,x_2,x_3,x_4,1/2-x_5,1/2-x_6,1/2-x_7)
\ee
The singular set of ${T^7}/\sigma$ contains eight copies of $T^4$. Now let us
consider the action on these of $\alpha$ and $\beta$ which define the 
manifold. Firstly $\beta$ exchanges these eight four-tori in four pairs of
two, leaving four independent elements. Thus, the singular set of
${T^7}/{(\sigma,\beta)}$ contains four copies of $T^4$ associated with 
$\sigma$. However, the fixed
points of $\sigma$ intersect those of $\alpha$, hence the singular set of
${T^7}/{(\alpha,\beta,\sigma)}$ contains four copies of ${T^4}/{Z_2}$ which
are associated with $\sigma$, where the $Z_2$ is the action of $\alpha$ on
the four-tori fixed by $\sigma$. It is easily seen that the action of $\alpha$
on these four-tori defines a $K3$ orbifold metric for each one. Finally, we
can say that the fixed set of $\sigma$ in ${CY}_{19}\times{S^1}$ is four
copies of $K3$. ${CY}_{19}\times{S^1}$ thus contains at least four $K3$
submanifolds, all of which are coassociative. 

From our discussion above it follows that the dimension
of the moduli space of each of these is $b^{2}_{+}(K3)=3$. It is therefore
true that the manifold ${CY}_{19}\times{S^1}$ is at least locally fibered by
$K3$ \cite{J2}. In fact it is not difficult to see along similar lines that
all of the seven manifolds used for $M$-theory compactification in this
paper admit local $K3$ fibrations.

Using the same example as we have just discussed, consider the following
isometry of ${CY}_{19}\times{S^1}$:
\be
\sigma(x_1,...x_7) = (1/2-x_1,1/2-x_2,-x_3,-x_4,x_5,x_6,x_7)
\ee
It is not too difficult to convince oneself that the fixed set of $\sigma$ in
${CY}_{19}\times{S^1}$ is four copies of $T^3$. It follows that these are all
associative submanifolds of the seven manifold. Now, because these
three tori are labelled by $(x_5,x_6,x_7)$, which are also coordinates of
the heterotic background, the above isometry also fixes precisely four 
three-tori of the heterotic background. Unfortunately little appears
to be known about the moduli space of such submanifolds and so we cannot
conclusively say that the six manifold admits a $T^3$ fibration. However
consistency with the dualities presented here and with the general picture
of fibering existing dual pairs to obtain more dual pairs \cite{vaf}
suggests the following:
\newline
(i) That the Joyce seven manifolds used for $M$-theory compactifcation in this
paper admit global $K3$ fibrations over some three manifold $Y$.
\newline
(ii) That the six manifolds, used to produce the heterotic duals to the
above $M$-theory compactifications, admit global $T^3$ fibrations over the
same three manifold $Y$.

We have thus far given examples of $M$-theory compactifications on Joyce
manifolds which admit local $K3$ fibrations. The dual heterotic 
compactifications were found to be on particular Calabi-Yau spaces, 
which we generically denote by $CY$. However, it
is also possible to inerpret other $M$-theory duals of these heterotic
compactifications, as compactifications of the eleven dimensional theory on
$CY{\times}{{S^1}/{Z_2}}$ \cite{Hor}. This is analogous to the two $M$-theory
duals of the heterotic string in seven dimensions, which we discussed in
section (1). 

It thus appears that if the fibration picture \cite{vaf} does hold
for the examples presented here, then even though we have constructed new
dual pairs with non-freely acting orbifolds of existing dual pairs (namely
heterotic/$M$-theory duality in seven dimensions), the adiabatic argument
is not violated. This is only true because the orbifolds to which we have
restricted our attention are precisely those which preserve (at least locally)
the fibration of the seven manifold. It is thus presumably true that if one
can find examples of orbifolds with higher order isometry groups which are
also of this type, then one may be able to construct more examples, some
of which may be phenomenologically appealing.

\section{Summary and Conclusions}
Summarizing the key points of this paper: 
For the conjectured dualities between the various string theories
and $M$-theory, 
the theories should yield identical massless spectra.
We began with the 11d-theory on a $K3{\times}T^3$ orbifold and the heterotic
string on $T^6$. We then orbifolded the seven manifold on the
$M$-theory side and resolved the singularites by blowing up, first
for an $N=2$ example and then for several $N=1$ examples. We presented an
ansatz mapping six of the $M$-theory coordinates to the six heterotic 
coordinates. Because of this ansatz, the action of the orbifold isometry 
group on the $M$-theory
geometry also specified the action on the heterotic geometry.
Thus all that remained to specify the heterotic background
was the embedding of the spin connection in the gauge connection. This was
derived by requiring modular invariance at the orbifold limit. Using
this ansatz we rederived previous results in the literature and produced
new dual pairs in four dimensions. It therefore appears
that this ansatz is consistent for producing dual pairs by orbifolding
whenever the non-freely acting elements of the orbifold group have
singularities which are of the $K3$ orbifold type. This is in accord with
the underlying structure of $K3$ fibrations, 
\cite{klemm,k,vaf}, because, as was demonstrated in
\cite{J2}, and as we showed in the last section ,
manifolds which are constructed with such orbifold elements admit
(at least locally) a $K3$ fibration. The duality is then expected to hold on
general grounds {\it if} the heterotic geometry is a $T^3$ fibration 
\cite{vaf}.

Moreover, we also gave further evidence that $M$-theory is consistent on
orbifolds by comparing the untwisted and twisted matter content in such
a theory with its heterotic dual. The untwisted matter contents for $N=2$
and $N=1$ dual pairs always agree for the dual pairs constructed
according to our ansatz. Specifically, if the four dimensional
$M$-theory background is of the form $CY{\times}S^1$, where $CY$ is
a ${Z_2}{\times}{Z_2}$ orbifold of $T^6$, 
then the dual heterotic background will,
according to our ansatz, necessarily be of the form ${{T^4}/{Z_2}}{\times}T^2$
. Both of these $N=2$ theories have an untwisted matter content of four
hypermultiplets. Similarly, $M$-theory on a Joyce $G_2$ orbifold of the form
${T^7}/{Z_2^3}$ will always contain an untwisted sector of seven $N=1$ moduli 
multiplets. Then, according to our ansatz, the dual heterotic background will
be of the form ${T^6}/{{Z_2}{\times}{Z_2}}$, and will also have seven $N=1$
moduli multiplets in the untwisted matter sector. What is even more compelling
is that we were able to show that for some simple choices of orbifold
gauge embeddings, the number of twisted sector multiplets agree also.

Finally we wish to add that this method of construcing dual pairs has
been successfully applied to the construction of $N=1$ dual pairs in
three dimensions \cite{ba}. In fact, using this construction we have 
constructed heterotic duals (on Joyce $G_2$ manifolds) for $M$-theory
compactifications on {\it all}
known Joyce manifolds of Spin(7) holonomy \cite{J3}.
This may shed some light on the mysterious
supersymmetric theory in twelve dimensions which appears to be required to
complete the duality picture \cite{Chris,F}. As noted in \cite{F}, these
theories may be an explicit realisation of the beautiful ideas of Witten
\cite{Cos} which may solve some of the long standing problems of
theoretical physics and possibly take duality towards reality.

{\bf Acknowledgements}.

The author is extremely indebted to Jerome
Gauntlett, Chris Hull,
Tomas Ortin, Bas Peeters, 
Wafic Sabra, Ashoke Sen and Steven Thomas for discussions.
The author would also like to thank PPARC, by whom this work is supported.

\end{document}